\documentclass[journal=aelccp,manuscript=letter,layout=twocolumn]{achemso}

\usepackage{lineno,hyperref}
\modulolinenumbers[5]

\usepackage[utf8]{inputenc} % allow utf-8 input
\usepackage[T1]{fontenc}    % use 8-bit T1 fonts
\usepackage{hyperref}       % hyperlinks
\usepackage{url}            % simple URL typesetting
\usepackage{booktabs}       % professional-quality tables
\usepackage{amsfonts}       % blackboard math symbols
\usepackage{nicefrac}       % compact symbols for 1/2, etc.
\usepackage{microtype}      % microtypography
\usepackage{lipsum}
\usepackage{color,soul}     % for comment hl
\usepackage{amsmath}        % for align % Used for the boldsymbol.
\usepackage{graphicx}       % use graphics
\usepackage{todonotes}
\usepackage{booktabs}
\usepackage{tabularx}

\usepackage{subfig} % use multiple figures

\usepackage{setspace}
\usepackage{subfig} % use multiple figures

\usepackage{nomencl}
\renewcommand{\nomname}
\makenomenclature

\usepackage[version=4]{mhchem}

\usepackage{color,soul}
\usepackage[]{xcolor}

\title{Physics-Guided Continual Learning for Predicting Emerging Aqueous Organic Redox Flow Battery Material Performance}
\author{Yucheng Fu}
\email{yucheng.fu@pnnl.gov}
\affiliation{Advanced Computing, Mathematics and Data Division,  Pacific Northwest National Laboratory,  Richland, WA 99354}
\author{Amanda Howard}
\affiliation{Advanced Computing, Mathematics and Data Division,  Pacific Northwest National Laboratory,  Richland, WA 99354}
\author{Chao Zeng}
\affiliation{Advanced Computing, Mathematics and Data Division,  Pacific Northwest National Laboratory,  Richland, WA 99354}
\author{Yunxiang Chen}
\affiliation{Advanced Computing, Mathematics and Data Division,  Pacific Northwest National Laboratory,  Richland, WA 99354}
\author{Peiyuan Gao}
\affiliation{Advanced Computing, Mathematics and Data Division,  Pacific Northwest National Laboratory,  Richland, WA 99354}
\author{Panos Stinis}
\affiliation{Advanced Computing, Mathematics and Data Division,  Pacific Northwest National Laboratory,  Richland, WA 99354}


\let\oldmaketitle\maketitle
\let\maketitle\relax

\begin{document}

\maketitle
\twocolumn[
\begin{@twocolumnfalse}
\oldmaketitle
\begin{abstract}
	Aqueous organic redox flow batteries (AORFBs) have gained popularity in renewable energy storage due to their low cost, environmental friendliness and scalability. The rapid discovery of aqueous soluble organic (ASO) redox-active materials necessitates efficient machine learning surrogates for predicting battery performance. The physics-guided continual learning (PGCL) method proposed in this study can incrementally learn data from new ASO electrolytes while addressing catastrophic forgetting issues in conventional machine learning. Using a ASO anolyte database with a thousand potential materials generated by a 780 $\text{cm}^2$ interdigitated cell model, PGCL incorporates AORFB physics to optimize the continual learning task formation and training process. This achieves higher efficiency and robustness compared to the non-physics-guided continual learning while retaining previously learned battery material knowledge. The trained PGCL demonstrates its capability in assessing emerging ASO materials within the established parameter space when evaluated with the dihydroxyphenazine isomers. 

\end{abstract}
\end{@twocolumnfalse}
]

%%%%%%%%%%%%%%%%%%%%%%%%%%%%
%%% Introduction %%%
%%%%%%%%%%%%%%%%%%%%%%%%%%%%
\section{Introduction}
As global energy consumption continues to grow and the urgency of transitioning to renewable energy sources intensifies\cite{koohi2020review,huggins2016energy}, flow batteries have carved a significant niche for themselves in the market, primarily because of their ability to provide reliable and efficient energy storage solutions\cite{ding2013vanadium, kamat2017redox,weber2011redox}. The evolution of redox flow batteries (RFBs) has transitioned from metal-centered redox-active materials to organic molecular systems due to concerns about cost, resource availability, and environmental impact\cite{wang2018high,brushett2020lifetime, winsberg2017redox}. Among these, the aqueous organic redox flow battery (AORFB) stands out as a compelling choice by combining the benefits of organic redox-active materials with a water-based electrolyte\cite{singh2019aqueous,wang2016redox}. The aqueous soluble organic (ASO) redox-active materials are favorable due to their high solubility, versatility, and fast kinetics\cite{hollas2018biomimetic, wei2017materials,wang2013recent}. With inherent flow battery advantages like scalability and safety, AORFBs hold significant promise for future energy storage solutions.

Machine learning (ML) algorithms, including data-driven and physics-informed neural network models,\cite{he2022physics,he2022enhanced, howard2022physics,li2022machine}, have significantly advanced RFB development. They enable characterization of molecular properties, reaction kinetics, and prediction of cell performance across various scales \cite{panapitiya2022evaluation,gao2021graphical, bao2020machine,wan2022machine,shah2023new, wan2021coupled, li2020cost}. Despite these advancements, predicting AORFB performance remains a challenging area due to the lack of well-curated ASO redox-active material datasets. To maintain state-of-the-art performance, both data-driven and physics-informed ML models need updates with the addition of new ASO materials. However, ML models cannot be simply retrained using only new materials because they suffer from catastrophic forgetting and exhibit poorer performance on the previous data\cite{goodfellow2013empirical,kirkpatrick2017overcoming}. This constant need for retraining on the whole dataset poses a significant challenge, especially considering the fast pace of new AORFB material development\cite{zhang2022emerging}. For instance, the recent SOMAS database reported more than 12,000 molecule candidates that could be used for AORFB \cite{gao2022somas}. To address this, continual learning (CL) has emerged as a promising approach, allowing models to update and expand using solely new data, without forgetting previous acquired knowledge \cite{hadsell2020embracing, PARISI201954}. Although CL has gained attention in energy storage\cite{MASCHLER2022513, zhao2022battery, chen2024lifelong,zhang2017lifelong}, its application in flow batteries, particularly AORFBs, has not been reported yet.

\begin{figure*}[h]
    \centering
\includegraphics[width=0.65\textheight]{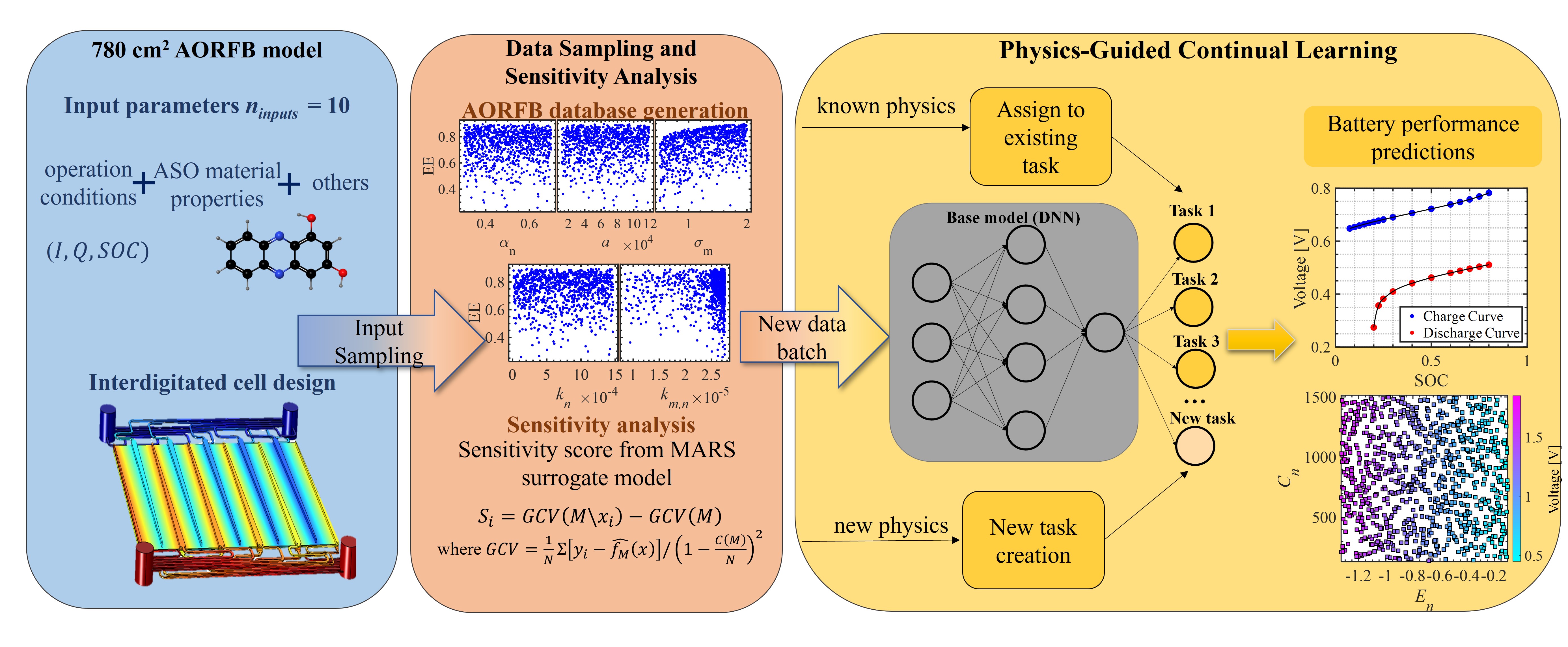}
    \caption{Illustration of the physics-guided continual learning framework.}
    \label{fig:PGCL_Framework}
\end{figure*}

In CL, a task is defined as a distinct aspect or a subset of data of a problem that the model is trained on\cite{hadsell2020embracing}. The main stream of CL algorithms are developed for classification problems, which have well-defined task boundaries. 
However, the exploration of CL applications in regression scenarios was not reported until the work by He and Sick in 2021 \cite{he2021clear}. Applying CL to regression tasks, such as predicting battery voltage or energy efficiency, introduce additional complexities. The high-dimensional nature of material properties further complicates the implementation of CL in AORFB systems, primarily due to the undefined strategies for determining task boundaries based on material and cell component properties and their impact on CL performance. Incorporating known physics into CL methods can provide significant advantages. In this context, we propose a Physics-Guided Continual Learning (PGCL) approach, specifically designed to forecast cell performance by leveraging the underlying physics of AORFB. The PGCL framework is depicted in Figure \ref{fig:PGCL_Framework}, focusing on its ability to integrate the intricate physics of AORFB materials. This integration facilitates a clear strategy and guidance for the selection and structuring of CL tasks, aiming to optimize performance. When new AORFB material is presented, the PGCL system evaluates whether to instantiate a new task or employ an existing one based on the physics underlying the new materials. This decision-making is achieved using the sensitivity analysis tool informed by the MARS surrogate model. If a new batch of data introduces new physics, the PGCL framework, informed by sensitivity analysis, can dynamically decide whether there's a need to introduce new tasks to accommodate fresh insights from the data or to leverage current CL tasks for subsequent predictions. Such an integration of physics into the CL algorithms mitigates the risk of spawning redundant tasks, ensuring predictions remain both swift and precise. Furthermore, to optimize the PGCL model's efficacy, we have conducted a thorough investigation into the relationship between AORFB material properties and overall battery performance. This research delved into how these variables influence CL's predictive speed and accuracy. Having tested and fine-tuned task sequences, numbers, and other parameters, our PGCL model has demonstrated its efficiency and robustness, especially when predicting AORFB performance in complex scenarios.

\begin{figure*}[h]
    \centering
\includegraphics[width=0.65\textheight]{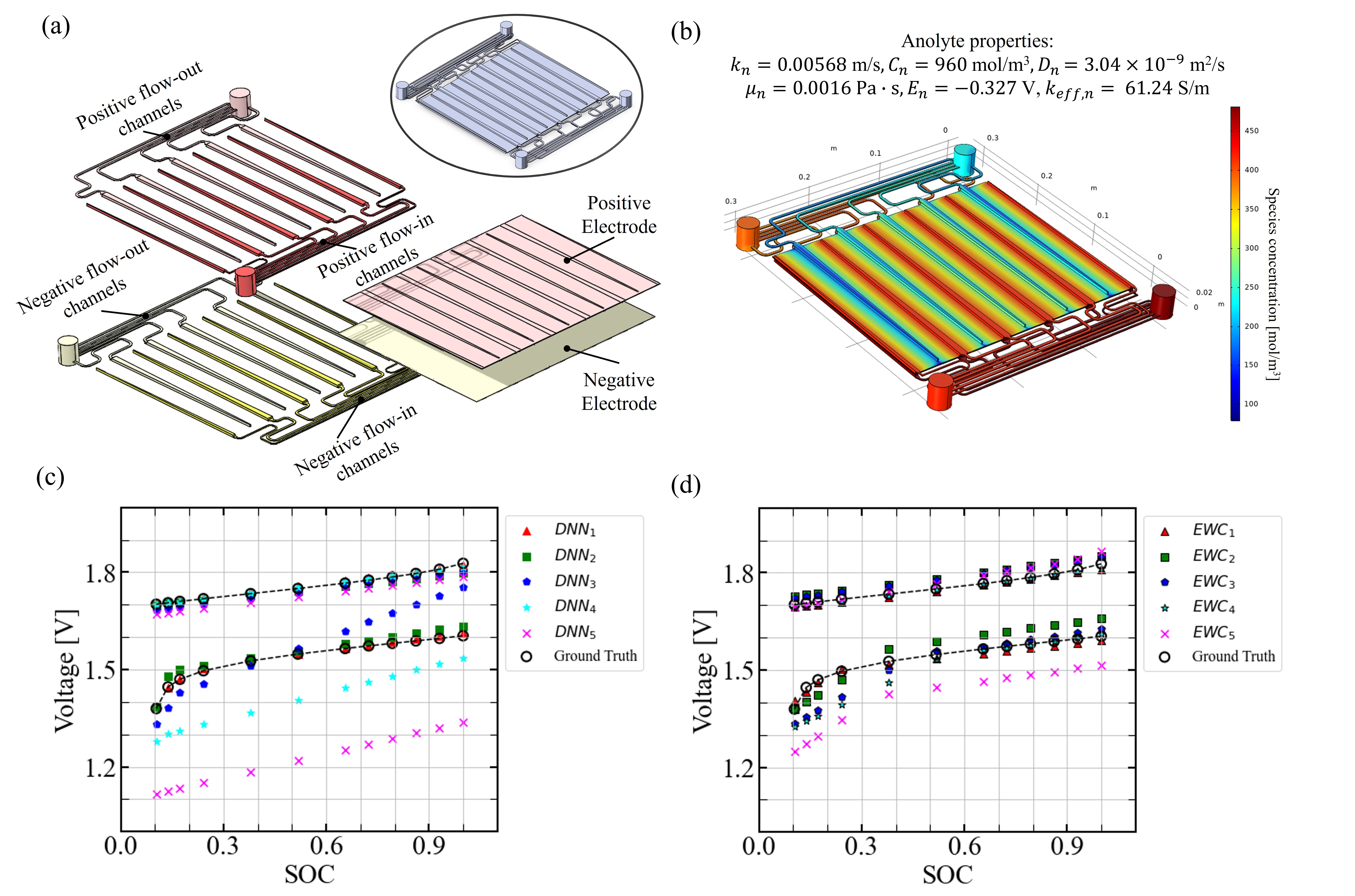}
    \caption{(a) Design of the 3D 780 $cm^2$ AORFB model and (b) redox-active species concentration distribution inside the flow channel with given anolyte properties. Comparison of the charge-discharge curve predictions using (c) a DNN without CL and (d) a DNN with CL (EWC is used here).}
    \label{fig:AORFB Model_Forgetting}
\end{figure*}

To assess the ASO material performance at practical operation conditions, a large physics-based 780 $cm^2$ interdigitated (ID) cell model is used for data generation as shown in  Figure \ref{fig:AORFB Model_Forgetting} (a). The model geometry follows the design reported by Reed et al. \cite{reed2016performance}, which includes the ID flow channels, electrodes, membrane and current collectors. In 3D models, the non-uniform distribution of species, current and flow distributions inside large-scale cells can be captured. Figure \ref{fig:AORFB Model_Forgetting} (b) shows an example of the redox-active species concentration distribution during discharge at State of Charge (SOC) of 0.5. The sampled anolyte material and cell properties are listed alongside the figure. As can be seen in the cell, the inlet flow channel has the highest species concentration, which is gradually reduced when going through the electrode and entering the flow-out channels.
The coupled physics of electrolyte hydrodynamics, species transport, electrochemical reactions have been incorporated in the model and more details are provided in Support Information S1. Using Latin Hypercube Sampling (LHS)\cite{anderson2009response, florian1992efficient}, a thousand ASO materials with various properties have been selected as input for the 780 $cm^2$ ID cell model. The charge-discharge curve and the energy efficiency (EE) are calculated for each sampled material, as detailed in Support Information S2. Materials that yield no meaningful physical results are omitted from consideration.

To demonstrate the issue of catastrophic forgetting, the battery voltage curve data were divided into five batches based on the material standard potential $E_n$. A Deep Neural Network (DNN) was sequentially trained on these data batches. After each training session, the resultant DNN, denoted as $DNN_i$, was employed to predict the charge-discharge curve of the specified ASO material, with an $E_n$ value that belongs to the first data batch. Initially, as depicted in the figure, the DNN predicts the charge-discharge curve with commendable accuracy. However, beginning with the model trained on the third data batch, a significant deviation from the ground truth is observed. Particularly after the fifth training batch, the maximum voltage prediction error observed at the discharge tail end reaches as high as 0.3 V.
Implementing a CL method, such as Elastic Weight Consolidation (EWC), mitigates this issue of catastrophic forgetting, as illustrated in Figure \ref{fig:AORFB Model_Forgetting} (d). With the EWC method applied, the prediction errors are confined within 0.1 V after training sequentially on four data batches and remain under 0.2 V thereafter. This illustrates the effectiveness of CL strategies in preserving learned knowledge and maintaining prediction accuracy throughout the sequential learning process for new and varying data sets.

To compare the CL performance with the traditional non-regulated DNN, two popular CL algorithms, Elastic Weight Consolidation (EWC)\cite{kirkpatrick2017overcoming}  and Learning without Forgetting (LwF)\cite{li2017learning}, are selected as the benchmark methods. As visualized in Figure S2 (Support Information), the CL methods demonstrate superior performance for remembering the previously trained ASO materials with errors reduced by up to 10 times compared to the regular DNN method for old tasks.

\section{Results and Discussions}\label{sec:ResultsDiscussions}
\subsection{Continual Learning Strategies with Redox Flow Battery Physics}\label{sec:TaskDivi}

\begin{figure*}[h]
    \centering
\includegraphics[width=0.7\textheight]{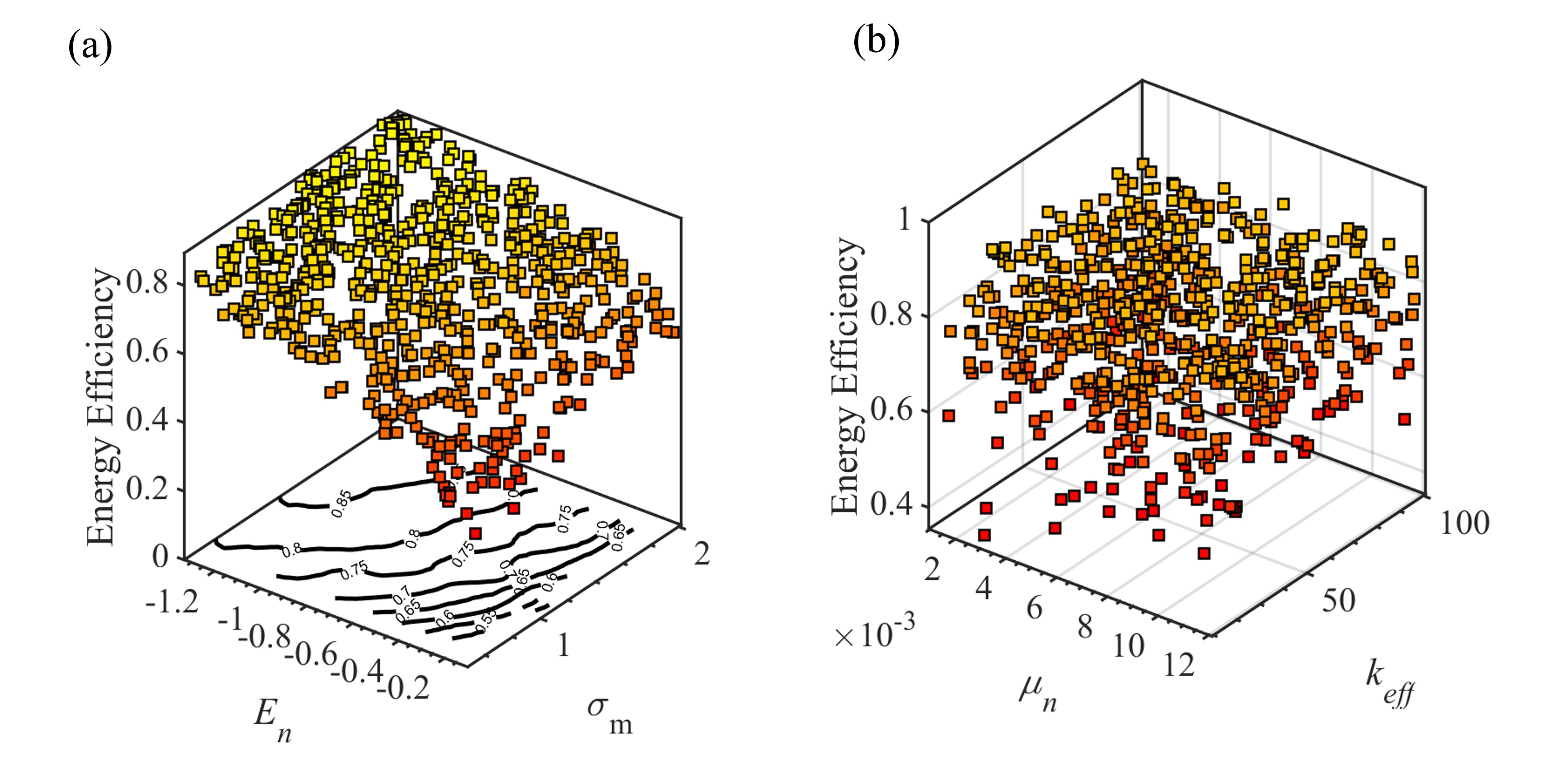}
    \caption{Visualization of the AORFB energy efficiency with respect to the anolyte parameters (a) $E_n$ and $\sigma_m$, (b) $\mu_n$ and $k_{eff}$.}
    \label{fig:Parameter_vs_EE}
\end{figure*}

To achieve optimal CL performance, it is important to incorporate AORFB battery physics into the CL task division strategies. We first explore how the selection of material and cell properties used for task partitioning can affect the CL algorithm performance. To better understand the relationship between the input parameters and the output EE, a sensitivity analysis using the MARS surrogate is conducted to identify the most sensitive anolyte and cell properties (Section S2, Support Information). Figure \ref{fig:Parameter_vs_EE} shows two 3D scattered plots visualizing the material properties vs. EE. The anolyte standard potential ($E_n$) and the membrane ionic conductivity ($\sigma_m$) show strong effects on the battery energy efficiencies.  With a decrease of $E_n$, the cell equilibrium potential would increase, resulting in a higher EE. For $\sigma_m$, an increase of its value reduces the system ohmic resistance, which results in EE improvement. At the corner with $E_n$ around -0.2 V and $\sigma_m$ at 0.5 [S/m], the EE sharply drops toward 0.2, indicating that the battery operates in extremely unfavorable conditions. On the other hand, parameters such as electrolyte viscosity ($\mu_n$) and ionic conductivity ($k_{eff}$) have a trivial effect on EE. 

\begin{figure*}[h]
    \centering
\includegraphics[width=0.65\textheight]{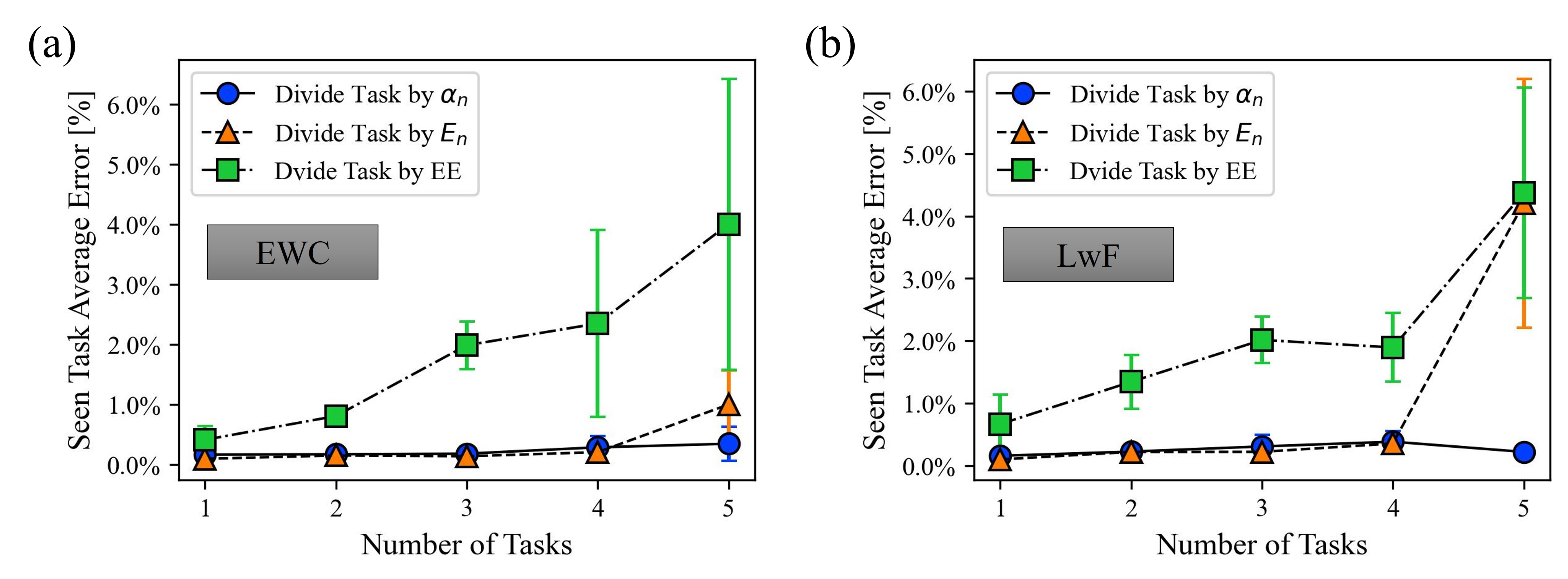}
    \caption{Comparison of how task division strategies influence the error of energy efficiency predictions using (a)EWC method and (b)LwF method. The strategies of dividing the CL tasks by the value of ASO material reaction transfer coefficient $\alpha_n$, standard potential $E_n$ and cell EE are tested respectively.} 
    \label{fig:EE_TaskDivision_Strategy}
\end{figure*}

With the knowledge gained from the sensitivity analysis, the EWC and LwF performances are evaluated on a set of tasks divided based on the range of $\alpha_n$, $E_n$ and $EE$, respectively (Figure \ref{fig:EE_TaskDivision_Strategy}). The number of tasks represents the current available tasks that have been used for training. The EE prediction error is determined by calculating the mean squared error (MSE) between the predicted EE values and the ground truth for the testing data of each task. This error is then averaged across all tasks seen so far, providing a cumulative measure of the model's predictive accuracy across the entire sequence of tasks. The error bars stand for the variation of the averaged error after 10 repetitions of the entire training process, with different random initial conditions for the CL network weights. As can be seen in the figure, when forming the tasks based on the range of the non-sensitive parameter $\alpha_n$, the seen task average error is fairly consistent with the increase of available tasks. If for the task division we use the more sensitive parameter $E_n$, both EWC and LwF show a slow increase of the mean error, which stays well below 1\% for the first four tasks. After providing the fifth task for training, the error observes a sharp change to around 4\% for both the EWC and LwF methods. For $E_n$, the first four tasks cover a potential range from -1.4 V to -0.26 V. Recall in Figure \ref{fig:Parameter_vs_EE}, the first four tasks ($E_n$ based division) cover an EE range from around 0.65 to 0.9. For the fifth task, even though $E_n$ ranges from -0.26 V to -0.1 V, the EE can vary from 0.65 to 0.25 due to the strong influence of this sensitivity parameter.  This wide $E_n$ range and the increased granularity of the fifth task cause a sharp increase of the CL prediction error. This observation is further confirmed if dividing the tasks based on the EE range directly. 

This example shows that the selection of task division parameter plays a crucial role in training and evaluating CL algorithms for AORFB systems. With the expansion of a sensitive parameter range, additional tasks would be necessary to learn the new physics of the materials. On the other hand, the CL algorithm performance has minimal variations if newly introduced tasks have only distinct non-sensitive parameter ranges. In those scenarios, the training of new tasks for CL can become unnecessary due to the increased time cost and marginal accuracy improvement. Additional tests on CL task number and sequence influence are further discussed in the Support Information S4. In short, these data division strategy tests highlight the importance of careful task division in an AORFB battery database, and the sensitivity analysis should be an important step to understand the AORFB physics when applying the CL method. 

\subsection{Physics-Guided CL Method}\label{sec:RealWorld}
\begin{figure*}[htbp]
    \centering
    \makebox[\textwidth]{\includegraphics[width=0.65\textheight]{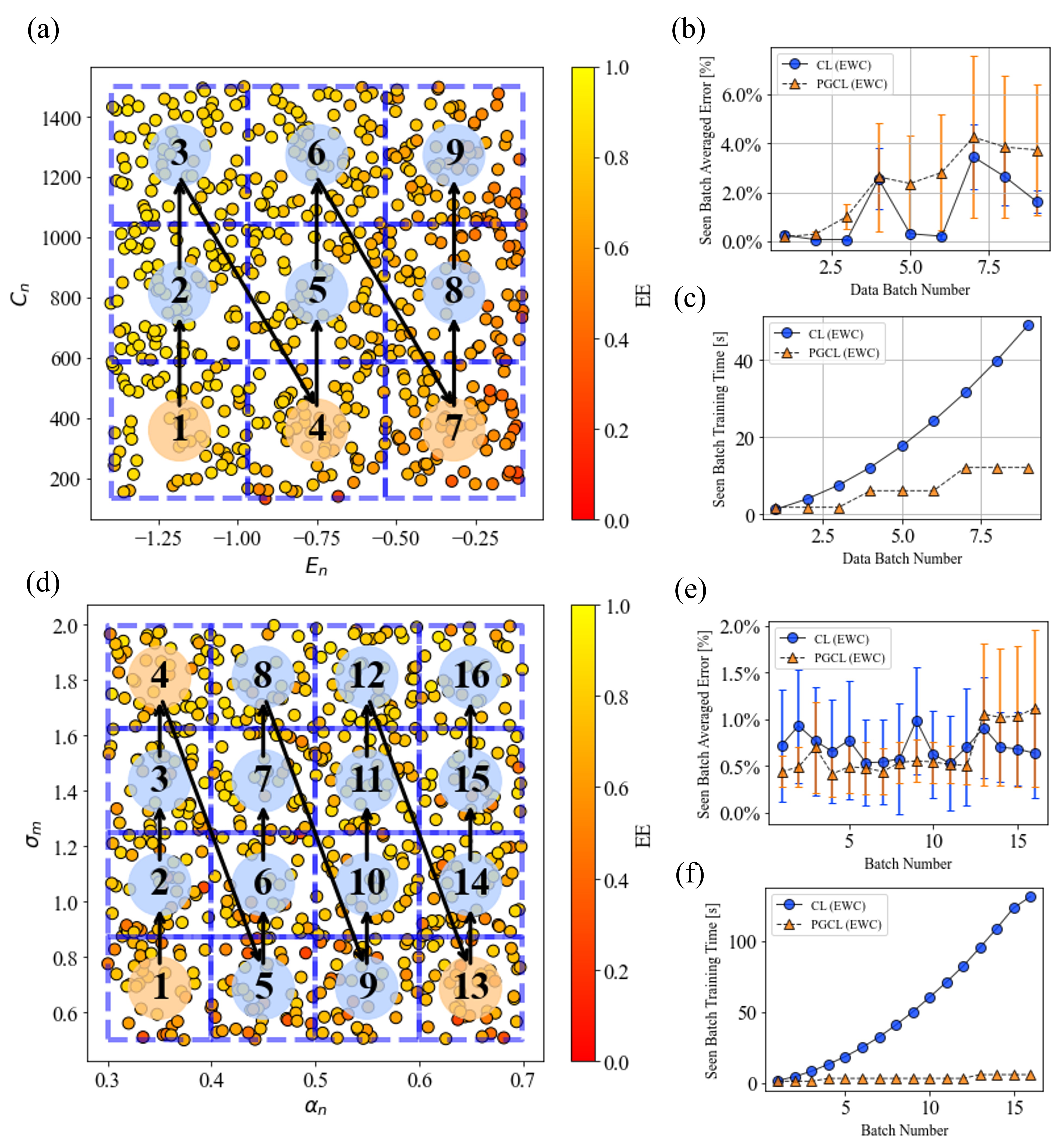}}
    \caption{Comparison of CL and PGCL using a sequentially introduced data batches with varying ASO material properties. The dataset is segmented based on the ranges of $C_n$ and $E_n$ into 9 distinct data batches, as illustrated in (a), and based on the ranges of $\sigma_m$ and $\alpha_n$ into 16 data batches, as shown in (d). Yellow circles indicate the specific data batches at which the PGCL algorithm initiates a new task during the training process. Panels (b) and (c) compare the accuracy and training time costs between traditional CL and PGCL for the first scenario in (a), while (e) and (f) provide comparisons for the second scenario in (d).}
    \label{fig:RealScenario}
\end{figure*}

In the context of redox-flow battery applications, the task formulation should not be confined to using a single parameter. New ASO materials often exhibit complex variations across multiple properties. Figure \ref{fig:RealScenario} shows two more complex scenarios with data batches divided based on two properties of the ASO material. In the first scenario, depicted in Figure \ref{fig:RealScenario}(a), the ASO materials are divided into nine batches based on the initial concentration $C_n$ and standard potential $E_n$. The second scenario in Figure  \ref{fig:RealScenario} (d) provides 16 data batches divided using the reaction transfer coefficient $\alpha_n$ and membrane ionic conductivity $\sigma_m$ values. The color coding of the scattered dots stands for the AORFB EE with corresponding anolyte material. During the training, the data batches are made available to the CL algorithm in the sequence marked by the black arrow and the number in each circle.

By taking into account the related physics of the AORFB cells, we proposed a PGCL method that can further improve the CL algorithm performance by optimizing the timing and strategies of new task creation and data grouping. The time cost, accuracy, and complexity of the regular CL and PGCL training methods are compared. For standard CL methods, each new data batch leads to the creation of a new task, resulting in nine tasks for the scenario demonstrated in Figure \ref{fig:RealScenario} (a). However, the PGCL optimizes this process by utilizing insights from AORFB material properties. When a new batch introduces changes only in non-sensitive properties, CL does not encounter new physics, and prediction errors remain stable despite additional tasks. Thus, PGCL groups new data into existing tasks with similar sensitive property ranges, avoiding unnecessary task creation. In the Figure \ref{fig:RealScenario}(a), the data batch marked with a yellow circle indicates the new task created for the PGCL, while those in blue were handled using existing tasks. This reduces the required task from nine to three with the PGCL. In the following comparison, the EWC architecture is used for the CL algorithms and each test is repeated 10 times. As shown in Figure \ref{fig:RealScenario}(b), the accuracy of the regular CL and PGCL are overall comparable across all the data batches with errors all below 4\%. For cases 2, 3, 5, 6, 8, and 9, CL performs slightly better due to additional tasks created for the data batches with the same $E_n$ range. However, this results in a much larger time expense. As shown in Figure \ref{fig:RealScenario}(c), the regular CL method training time increases monotonically with each added data batch. After training on all nine data batches, it takes 48 s to finish the whole training process. On the other hand, the PGCL takes less than 10 s to finish the training process while maintaining a similar level of accuracy. As for the 2, 3, 5, 6, 8, and 9 batches, the PGCL does not require additional training time for those data batches. The existing task can be used directly for EE prediction by comparing the $E_n$ range of the new dataset. A second test scenario, shown in Figure \ref{fig:RealScenario}(d), divides ASO materials into 16 batches based on two non-sensitive parameters, $\sigma_m$ and $\alpha_n$, with respect to energy efficiency (EE). Recognizing that changes in these non-sensitive parameters minimally impact CL predictions, PGCL selectively creates new tasks only for batches 1, 4, and 13, which covers the extreme range of $\sigma_m$ and $\alpha_n$. This significantly saves time compared to 16 tasks used for CL method. The EE prediction error is maintained at the range from around $0.5\%$ to $1\%$, affirming that non-sensitive parameter variations do not drastically affect performance. Moreover, with the physics learned from the AORFB system, PGCL can be customized based on different priorities. If accuracy is a priority, PGCL can be configured to generate a new task for each data batch provided. This would essentially result in the same accuracy predicted by the CL method.

\subsection{Prediction for New AORFB Materials}\label{sec:unseen materials}
To better evaluate the capability of the proposed PGCL method for predicting new AORFB materials, we assess its performance using the dihydroxyphenazine (DHP) isomers as the unseen anolyte materials. The selection of five DHP molecular properties, previously documented by Wellala \cite{wellala2021decomposition}, are utilized in the $780 cm^2$ cell model to calculate their cell EE as the ground truth. Significantly, our PGCL model was exclusively trained using data from the LHS-sampled AORFB material database.

\begin{figure*}[htbp]
    \centering
    \makebox[\textwidth]{\includegraphics[width=1\textwidth]{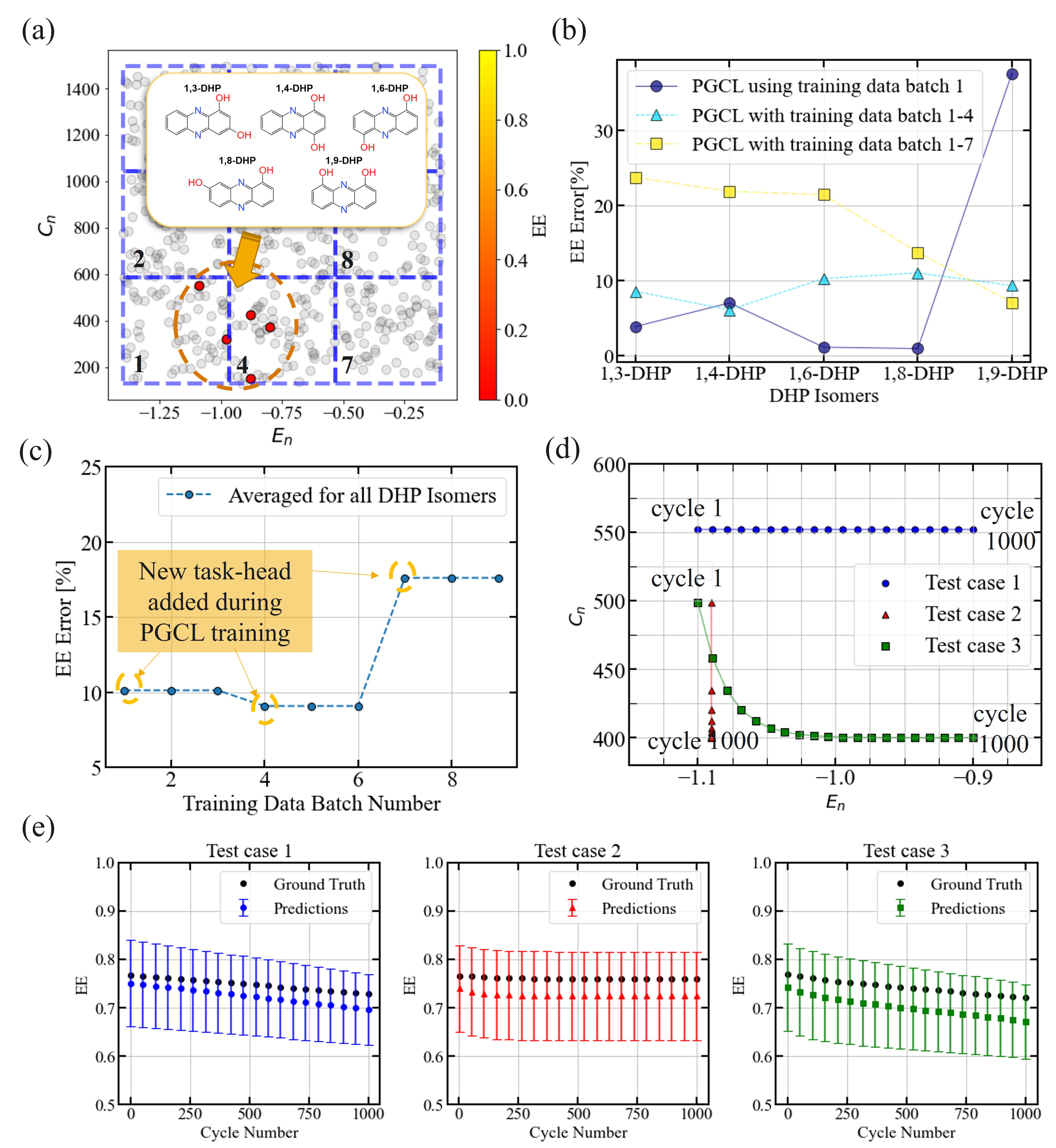}}
    \caption{Prediction of unseen DHP redox-active species performance. (a) The distribution of DHP isomer concentration and standard potential $E_n$ overlaid on the AORFB database. (b) The EE prediction accuracy of each DHP isomer with progress in PGCL training. (c) Averaged error of all DHP isomer predictions at different training data batches. (d) Three test cases with hypothetical 1,8-DHP material property degradation. (e) Comparison of the PGCL predicted EE against ground truth for three test cases over 1000 cycles.}
    \label{fig:DHP_Predictions}
\end{figure*}

Figure \ref{fig:DHP_Predictions} (b) illustrates the PGCL's prediction accuracy for each DHP isomer. The PGCL is trained following the process proposed in Figure \ref{fig:RealScenario} (a) by dividing the AORFB database into nine batches. As shown in the figure, the EE predictions of most DHP isomers fall within a error of less than $25\%$, which holds true across all tasks that the PGCL has encountered during training. The 1,9-DHP EE prediction shows more substantial fluctuation with the progress of PGCL training. This particular isomer, with a lower concentration of $C_n = 152 mol/m^3$, is situated at the edge of the ASO database concentration range where the EE sensitivity is heightened. Further insights can be drawn from Figure \ref{fig:DHP_Predictions} (c), which plots the MSE for all DHP cells as the PGCL model progresses through the nine training data batches. The fluctuations in error are observed during the first, fourth, and seventh batches. This is attributed to PGCL's function logic, which triggers the creation of new task heads only when additional physics is introduced by the feeding data batch. For practical applications, if no new materials emerge within the parameter domain delineated by batches 7-9, PGCL training could optimally halt at batch 6. Such a cutoff would ensure the efficiency of the PGCL, particularly for predicting DHP isomer performance, while also illustrating the benefits of continual learning, which mitigates catastrophic forgetting by dynamically incorporating new data.

The PGCL methodology is developed for encompassing a range of anolyte material properties, enabling it to predict cell performance and degradation across multiple cycles. Currently, characterizing the degradation mechanisms for ASO redox-active materials is still an challenge task. The ASO materials stability is affected by its chemical structures, operational conditions and often the interactions with other cell components, such as membranes, electrodes, and electrolytes additives, which the PGCL approach cannot directly predict \cite{wedege2016organic,kwabi2020electrolyte}. Nevertheless, once the material degradation pathways are determined, PGCL can effectively forecast the cell performance over multiple cycles. As an illustrative case, we utilized 1,8-DHP as a baseline to predict the cell EE under three hypothetical material property degradation scenarios over 1000 cycles. The first test case hypothesizes that 1,8-DHP decomposition predominantly drives performance degradation, as reflected in $C_n$ changes over time. The second scenario speculates alterations in the $E_n$ of 1,8-DHP across the cycle lifespan. The third scenario combines both $E_n$ and $C_n$ changes to assess their cumulative effect. Ground truth energy efficiency (EE) are calculated for these scenarios over 1000 cycles with the physics-based 780 $cm^2$ cell model. Figure \ref{fig:DHP_Predictions} (e) presents the comparison of predicted EEs against the ground truth for all three test cases. The predictions align with the actual trends, maintaining an absolute error within $10\%$. As machine learning algorithms evolve to characterize new ASO material degradation pathways, PGCL can be integrated as a robust predictive tool for cell performance assessment over extended cycling periods.

\section{Conclusions}\label{sec:Conclusions}
In summary, the PGCL algorithm has been developed to dynamically learn and predict the AORFB performance with the emerging new ASO redox-active materials. PGCL effectively addresses the catastrophic forgetting issues often encountered with traditional, non-regulated Deep Neural Network (DNN) methods. 
By incorporating physical principles, PGCL enhances the prediction accuracy for new AORFB materials, highlighting its adaptability and efficiency in dynamic material discovery scenarios, in comparison to the conventional CL methods. The key insights gained by the PGCL for the AORFB system can be summarized into the following points:
\begin{itemize}
\item The standard potential, membrane conductivity, and species solubility are important factors dictating AORFB performance.
\item Task division strategies significantly impact CL algorithm performance, and task creation should occur only with expansion of ASO materials containing sensitive properties.
\item The total number of CL tasks and the sequence should be determined based on the data granularity to reduce the AORFB EE prediction errors.
\item PGCL not only streamlines the training process but also offers a structured approach for task division based on the physical properties of ASO materials.
\end{itemize}
In essence,  PGCL highlights the importance of integrating physics and understanding how material properties influence cell output in determining task division strategies, sensitivity parameters, and task sequences for optimizing CL algorithms in AORFB systems. Currently, PGCL employs a data-driven DNN as the regression surrogate model for task training. The performance of PGCL could be further enhanced by integrating a more advanced physics-informed surrogate or molecular description to better guide the realization of practical molecules in the ASO material data parameter space.

The principles and procedures outlined in our PGCL framework extend beyond AORFB systems to encompass a wide range of energy storage technologies, including lithium-ion batteries, solid-state batteries, and more. Its core strength lies in its adaptability and efficiency in learning from new materials, eliminating the need for repetitive training on the existing data. The strategies for task creation and division developed in this study, informed by PGCL, can be adapted to other regression surrogates for creating CL models aimed at predicting various cell performance metrics. This attribute is particularly valuable as the field of energy storage materials undergoes rapid evolution and diversification. By incorporating the physics of the studied system into CL, our PGCL variant can  facilitate the swift development of efficient and cost-effective energy storage solutions.

%%%%%%%%%%%%%%%%%%%%%%%%%%%%%%%%%%%%%%%%%%%%%%%%%%%%
%%%     End of body of article     %%%
%%%%%%%%%%%%%%%%%%%%%%%%%%%%%%%%%%%%%%%%%%%%%%%%%%%%
\begin{acknowledgement}.
This research was supported by the Energy Storage Materials Initiative (ESMI) under the Laboratory Directed Research
and Development (LDRD) program at Pacific Northwest National Laboratory (PNNL).  PNNL is a multi-program national laboratory operated for the U.S. Department of Energy (DOE) by Battelle Memorial Institute under Contract No. DE-AC05-76RL01830. \\
Additionally, the raw data and code that support the findings of this study are available from the corresponding author upon request.
\end{acknowledgement}

\bibliography{references}
\end{document}